# Inferring exemplar discriminability in brain representations

Hamed Nili[1], Alexander Walther, Arjen Alink[2], Nikolaus Kriegeskorte[3]

1    Wellcome Centre for Integrative Neuroimaging, University of Oxford
2    University Medical Center Hamburg-Eppendorf
3    Zuckerman institute, Columbia University

## Abstract

Representational distinctions within categories are important in all perceptual modalities and also in cognitive and motor representations. Recent pattern-information studies of brain activity have used condition-rich designs to sample the stimulus space more densely. To test whether brain response patterns discriminate among a set of stimuli (e.g. exemplars within a category) with good sensitivity, we can pool statistical evidence over all pairwise comparisons. Here we describe a wide range of statistical tests of exemplar discriminability and assess the validity (specificity) and power (sensitivity) of each test. The tests include previously used and novel, parametric and nonparametric tests, which treat subject as a random or fixed effect, and are based on different dissimilarity measures, different test statistics, and different inference procedures. We use simulated and real data to determine which tests are valid and which are most sensitive. A popular test statistic reflecting exemplar information is the exemplar discriminability index (EDI), which is defined as the average of the pattern dissimilarity estimates between different exemplars minus the average of the pattern dissimilarity estimates between repetitions of identical exemplars. The popular across-subject t test of the EDI (typically using correlation distance as the pattern dissimilarity measure) requires the assumption that the EDI is 0-mean normal under H0. Although this assumption is not strictly true, our simulations suggest that the test controls the false-positives rate at the nominal level, and is thus valid, in practice. However, test statistics based on average Mahalanobis distances or average linear-discriminant t values (both accounting for the multivariate error covariance among responses) are substantially more powerful for both random- and fixed-effects inference. Unlike average cross-validated distances, the EDI is sensitive to differences between the distributions associated with different exemplars (e.g. greater variability for some exemplars than for others), which complicates its interpretation. We suggest preferred procedures for safely and sensitively detecting subtle pattern differences between exemplars.

# 1. Introduction

Brain representations are increasingly investigated with pattern-information analyses of data acquired with brain imaging or neuronal recording techniques (Haxby et al., 2001; Hung et al., 2005; Kamitani and Tong, 2005; Kriegeskorte et al., 2006; Norman et al., 2006; Kiani et al., 2007; Kriegeskorte et al., 2008b; Mur et al., 2009; Kriegeskorte and Kievit, 2013; Kriegeskorte & Kreiman, 2011; Hebart and Baker, 2018, Jun et al, 2017, Kleinfeld et al., 2019). These analyses seek to quantify different types of information present in the brain-activity patterns. Information carried by brain response-patterns can be explored at different levels. Two common levels are the category and the exemplar level. Category information has previously been studied using regional average activation (e.g. Kanwisher et al., 1997) and pattern-decoding approaches (e.g. Haxby et al., 2001). Pattern-classifier decoding lends itself naturally to investigating category information with the category labels being decoded from the brain-activity patterns. Some studies have estimated a unique response pattern for each individual stimulus and investigated not only category information, but also within-category exemplar information (Mitchell et al., 2008; Kriegeskorte et al., 2008b). Exemplar information is present to the degree that different exemplars elicit distinct representational patterns. Within-category effects can be subtle, thus powerful tests are needed to detect them. Here we use the term "exemplar discriminability" generically to denote the average discriminability across all pairs among a set of experimental conditions.

Studying representational distinctions within categories can arise at different contexts. For example, in studies of visual face representations, it is important to quantify to what extent a face region distinctly represents individual faces (Kriegeskorte et al. 2007; Nestor et al., 2011; Anzellotti et al., 2013). Classifier decoding can be used to test for exemplar information. For example, a classifier can be trained to distinguish two exemplars within a category. For power, we would then like to have many repetitions of each stimulus. This would suggest repeating a small number of exemplars many times in the experiment (e.g. Kriegeskorte et al. 2007). It is also desirable, however, to sample the category with many different exemplars in order to get a richer description of its underlying representation. There is a tradeoff between the number of stimuli and the number of repetitions of each stimulus, because the total time available for measuring brain activity in a subject is usually limited. If we have many different exemplars (i.e. a condition-rich design), we can typically repeat each stimulus only a few times. This severely limits our power to detect the discriminability of a given pair of exemplars. In the case of condition-rich designs, it is therefore desirable to pool the evidence across many pairs of exemplars. This is achieved by a summary statistic that combines the evidence of discriminability across all pairs of exemplars.

A popular summary statistic reflecting exemplar discriminability among a set of experimental conditions from the same category is what we denote as the "exemplar discriminability index" (EDI; e.g. Sayres and Grill-Spector, 2008; Schwarzlose et al., 2008; Chan et al., 2010; Kravitz et al., 2010, Lee et al., 2012; Liu et al., 2013). The EDI is defined as the average between-exemplar dissimilarity estimate minus the average within-exemplar dissimilarity estimate (Fig. 1). A within-exemplar dissimilarity is a dissimilarity between independent measurements of the activity pattern elicited by repeated presentations of the same stimulus. The average within-exemplar dissimilarity estimate thus reflects the noise in the measurements. Subtracting it is essential when the dissimilarity estimate (i.e. pattern-distance estimates) is positively biased. The correlation distance, for example, which was used in the cited studies, is non-negative by definition, and therefore positively biased (although correlation coefficients are between -1 and 1, correlation distance, defined as 1 minus the correlation coeffiecient, is between 0 and 2). Subtracting the average within-exemplar dissimilarity removes the bias and enables inference. Typically, the EDI is computed for each subject and tested at the group level using a one-sided *t* test, treating subject as a random effect. It must be noted that the previous studies that used EDI do not use this nomenclature. However, we consistently use 'EDI' for any summary statistic computed in the way that is described above, i.e. any measure of exemplar discrimibaility that is computed by subtracting the within- from the between-exemplar dissimilarities.

The cited studies using this approach rely on splitting the data into two independent halves (*e.g.* odd and even runs in fMRI measurements) and comparing response patterns between the two halves. The between-halves dissimilarities are assembled in a split-data representational dissimilarity matrix (*sdRDM*), which is indexed vertically by response patterns estimated from data set 1 and horizontally by response patterns estimated from data set 2. Each entry of the sdRDM contains one dissimilarity between two response-pattern estimates spanning the two data sets. The order of the exemplars is the same horizontally and vertically, so the diagonal of the sdRDM contains the within-exemplar pattern dissimilarity estimates (Fig. 1A).

The within-data-set pattern dissimilarities are not used for either within- or between-exemplar pattern comparisons. This is important because patterns measured closer in time tend to be more similar due to measurement artefacts (Henriksson et al., 2014, Alink et al., 2015). Note that for the popular correlation distance (1 - Pearson r), the difference of distances is equal to the negative of the difference of correlations: d1 – d2 = (1-r1) - (1-r2) = r2 - r1. For a consistent comparison with other measures of pattern dissimilarity, however, we use the correlation *distance* here.

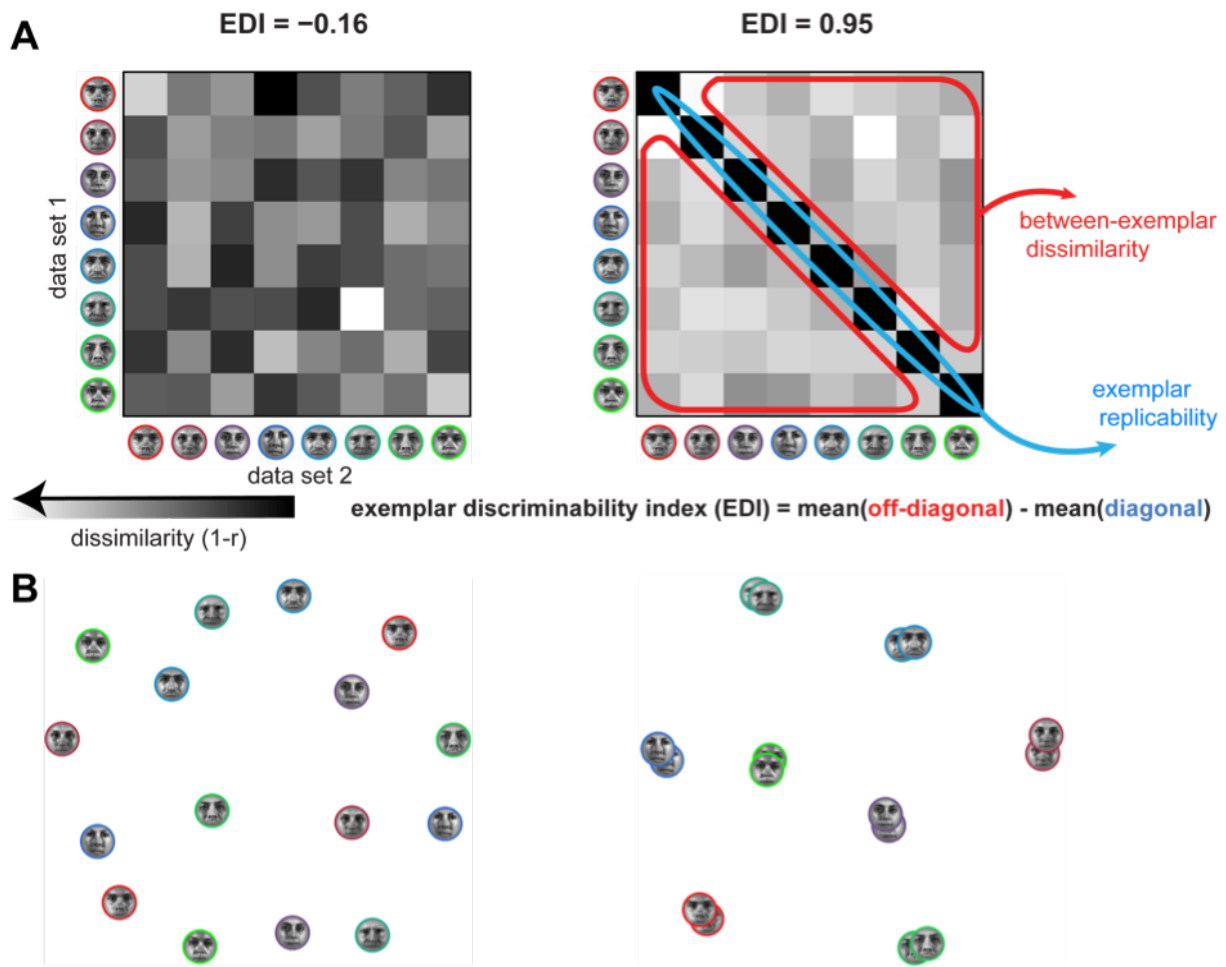

**Figure 1: Definition of the exemplar discriminability index (EDI). (A)** The exemplar discriminability index (EDI) is defined as the difference of the average between-exemplar and average within-exemplar distance. Split-data RDMs (sdRDMs) are shown for two scenarios. In the scenario on the left, the face exemplars are not discriminable and the EDI is insignificant (H0 simulation). In the scenario on the right, the EDI is large and exemplars are discriminable (H1 simulation). **(B)** Applying multi-dimensional scaling (MDS; Kruskal and Wish, 1978; Torgerson, 1958) allows simultaneous visualisation of the representational geometries from both data sets (H0 on the left, H1 on the right). Exemplars are colour-coded and there are two pattern estimates for each exemplar. MDS gives a low dimensional (2D here) arrangement, wherein the distances between projections approximate the original distances in the high-dimensional pattern space. Here, to obtain MDS plots, we consider a distance matrix composed of comparisons between any two patterns (aggregated from both datasets A and B). When the EDI is significantly positive (right), repetitions of the same exemplar yield patterns that are more similar to each other than presentations of two different exemplars.

The EDI *t* test has some caveats. First, this approach only allows testing exemplar information with subject as a random effect. Single-subject or group-level inference with subject as fixed effect cannot be accommodated in this approach. Second, it is possible that the assumptions of the test are not met. A *t* test requires that the data are normally distributed under $H_0$. The EDI is the difference of two average dissimilarities. For non-negative dissimilarities like the Euclidean distance, the distributions of the within- and the between-exemplar dissimilarities are skewed (limited by 0 on the left, unlimited on the right). Under the null hypothesis, the within- and between-exemplar dissimilarities are all samples from the same distribution (i.e.

for a given exemplar, the effect of changing the exemplar is not different from the effect of having another measurement for that exemplar). The expected value of the mean of the diagonal (within-exemplar) entries of the sdRDM is therefore equal to the expected value of the mean of the off-diagonal (between-exemplar) entries. The expected value of the difference between those means, i.e. the expectation of the EDI, is therefore also zero under $H_0$. However there are more between-exemplar than within-exemplar dissimilarities. For $N$ exemplars, the sdRDM has $N^2$ entries, so there are $N$ diagonal entries (within-exemplars) and $N*(N-1)$ off-diagonal entries (between-exemplars). Under $H_0$, the distribution of the average between-exemplar dissimilarities will therefore be narrower than that of the average within-exemplar dissimilarities. The EDI, thus, is a difference between two random variables, which have different variances and skewed distributions. The null distribution therefore need not be symmetric and can be non-Gaussian. The $t$ test, therefore, is technically invalid as a test of the EDI.

Fig. 2 illustrates different scenarios for the distributions of the diagonal and off-diagonal means under $H_0$. The distributions can be symmetric or skewed; furthermore they could have the same shape (implying also the same width) or different shapes (e.g. different widths). Note that a difference between two random variables is symmetrically distributed about 0 if either (a) the variables are identically distributed (can be skewed in this case) or (b) each of the variables is symmetrically distributed about the same expected value (they can have different shapes, e.g. different variances, in this case). However, the diagonal and off-diagonal means are both skewed and non-identical (different variances), so the EDI is not symmetrically distributed under $H_0$ thus not normally distributed. Therefore the assumptions of the $t$ test are not strictly met.

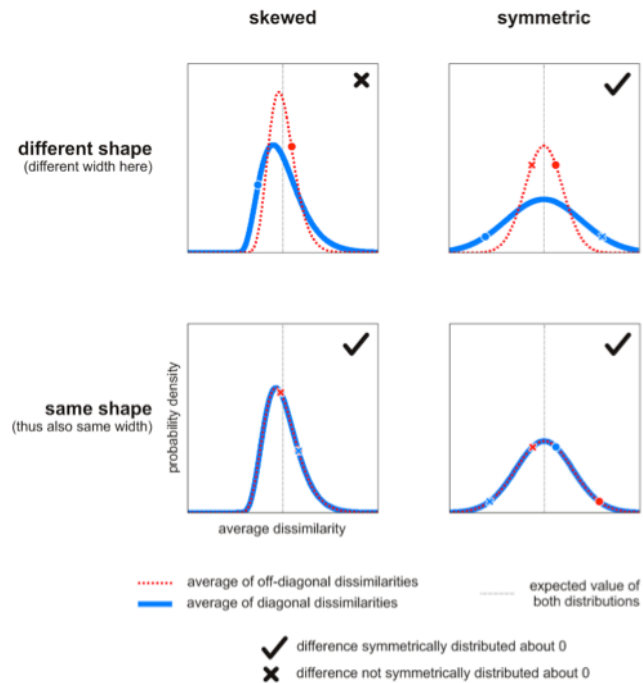

**Figure 2: The EDI is not strictly symmetrically distributed under H0.** EDI is the difference between the average diagonal entries (blue) and the average off-diagonal entries (red). The distributions of these averages can be skewed or non-skewed and have the same or different shapes. Note that under H0, the two distributions have the same expected value (gray vertical line). When the two distributions are both symmetric (right panels) or when they are both of the same shape (lower panels), their difference will be symmetrically distributed about 0. To see this, consider the fact that for any two points (one from each distribution, e.g. blue X and red X), another equally probable two points exist (blue O and red O) such that their difference has the same absolute value and the opposite sign. However, the distribution can be asymmetric (and thus non-Gaussian) when the two distributions are both skewed and have different shapes (upper left panel).

Testing hypothesis about representational geometries in RSA is achieved by comparing estimated representational dissimilarity matrices to hypothesized ones (e.g. obtained from computational models of information processing in the brain). The model-based RSA comparison quantifies the extent to which the two similarity structures (measured and hypothesized) conform. One important characteristic of representational geometries is the overall distinctness of the experimental conditions. In this paper, we explore different ways of quantifying and testing the overall decodability, which can be seen as model-free RSA: if the experimental conditions are discrimnable, they can be arranged in many different structures in the representational space. Note that using the term "exemplar" is only motivated by previous studies in vision scienecs and EDI can be used to quantify the overall discriminability of experimental condition in any experiment. Importantly, it can be used in condition-rich designs and does not require many repetitions of experimental conditions.

This paper has three aims. First, we assess the practical validity of the popular EDI *t* test by simulation. Second, we introduce a number of alternative tests that are valid and, unlike the

EDI *t* test, enable single-subject and group-level fixed-effect inference, along with random-effects inference. Third, we compare all these tests in terms of their power to detect exemplar discriminability in real data from functional magnetic resonance imaging (fMRI). The simulations show that the theoretical violations of the assumptions of the EDI *t* test are minimal in practice and hence the test is generally valid. However, exemplar discriminability tests based on dissimilarity measures that account for multivariate error covariance (Mahalanobis distance and linear discriminant *t* values; Kriegeskorte et al., 2006; Kriegeskorte et al., 2007; Nili et al.; 2014; Walther et al., 2016) are substantially more powerful and enable inference for single subjects. We also explain how EDI tests that are based on average cross-validated Mahalanobis distances are different from tests that are based on non-crossvalidated distances.

## 2. Material and methods

### 2.1 fMRI experiment

17 participants (10 female, age range 20-38) underwent four functional runs of scanning in two separate scanning sessions. In each run, participants were presented with 24 images of real-world objects belonging to two categories with 12 objects in each. Categories were changed from run to run towards ever more fine-grained categorical distinctions: animate/inanimate, faces/bodies, animal faces/human faces, and male faces/female faces. Participants were instructed to either categorize a stimulus based on the one previously shown (session one) or to complete a visual fixation task (session two). Each session also included two runs during which participants viewed retinotopic mapping stimuli (for details see Alink et al., 2013) and images depicting faces, houses, objects and scrambled objects. The analysis of brain responses to these stimuli allowed us to define regions of interest for FFA, PPA, LOC and early visual areas V1-3. Functional EPI images covering the entire brain were acquired on a 3T Siemens Trio scanner using a 32-channel head coil (32 slices, resolution = 3mm isotropic, inter slice gap =0.75mm, TR = 2000ms). For each participant we also obtained a high-resolution (1mm isotropic) T1-weighted anatomical image using a Siemens MPRAGE sequence.

All participants gave their informed consent after being introduced to the experimental procedure in accordance with the Declaration of Helsinki. The experimental procedure has been approved by the Cambridge Psychology Research Ethics Committee (ethics reference number: CPREC 2010.52)

## 2.2 Tests statistics for measuring exemplar discriminability

### 2.2.1 Exemplar Discriminability Index (EDI)

Representational similarity analysis (RSA, Kriegeskorte et al., 2008a), characterizes the representations of a brain region by a representational dissimilarity matrix (RDM). An RDM is a distance matrix composed of distances between response patterns corresponding to all pairs of experimental conditions. In cases where the responses to the same exemplars are measured in two independent data sets, a similar approach can be taken and the response patterns can be compared in a split-data RDM (sdRDM). Rows and columns of an sdRDM are indexed by the exemplars in the same order. Entries of this type of RDM are comparisons between responses to the same or different exemplars in two different data-sets.

Comparisons between the responses to two measurements of the same exemplar give the diagonal entries and correspond to the within-exemplar dissimilarities. Comparisons between the responses to two different exemplars give the off-diagonal entries, which correspond to the between-exemplar dissimilarities. For N exemplars, there would be N diagonal and N*(N-1) off-diagonal entries. The exemplar discriminability index (EDI) is calculated from an sdRDM by subtracting the two averages (Fig. 1A).

The EDI is the difference of the average between-exemplar and the average within-exemplar dissimilarities. Pattern dissimilarity (e.g. elements of the sdRDM) can be measured using various distance measures. In this paper, we explore the following measures of pattern dissimilarity:

- *Euclidean distance*: For two vectors **a** and **b**, the Euclidean distance is the $L^2$ norm of the difference vector **a**-**b**. Geometrically, it corresponds to the length of the vector that connects the two vectors (**a** and **b**) to each other.
- *Pearson correlation distance*: The Pearson correlation distance for two vectors **a** and **b** is equal to 1 minus the Pearson correlation coefficient between them. The correlation distance has a geometrical interpretation: It is one minus the cosine of the angle between the mean-centered vectors of **a** and **b** (e.g. voxel-mean centered transformations of **a** and **b**).
- *Mahalanobis distance*: The Mahalanobis distance is the Euclidean distance between the two vectors after multivariate noise normalisation. Multivariate noise normalisation is a transformation that renders the noise covariance matrix between the response channels identity. In this transformation, the response patterns are normalised through scaling with the inverse square root of the error-covariance. Therefore, if we have a response matrix, **B** (Response matrix is a matrix of the distributed responses to all exemplars, with each row containing the response to one exemplar in all response channels. The size of this matrix will be number of exemplars by number of response channels), we can noise-normalise it like so:

Eq. 1 
$$\mathbf{B}^* = \mathbf{B}\hat{\Sigma}^{-\frac{1}{2}}$$

where **B** is the original response matrix, $\hat{\Sigma}$ is the estimated error variance-covariance matrix (square matrix with number of rows and columns equal to the number of response channels, e.g. voxels), and $\mathbf{B}^*$ is the response matrix formed from the response patterns after multivariate noise normalisation (number of exemplars by number of response channels). If the number of response channels greatly outnumbers the number of exemplars, $\hat{\Sigma}$ will be rank-deficient and hence non-invertible. To ensure invertibility, we regularize the sample covariance estimate with optimal shrinkage (Ledoit & Wolf, 2004). This method shrinks the error covariance matrix towards the (invertible) diagonal matrix using an optimally weighted linear combination of the sample covariance matrix and the diagonal covariance matrix estimate.

The optimal shrinkage minimizes the expected quadratic loss of the resultant covariance estimate.

### 2.2.2 Average LD-*t/C*

Linear discriminant analysis (LDA) could also be used to estimate the discriminability values between pairs of exemplars. Similar to EDI, this method also relies on having two independent measurements (i.e. two datasets) for the same set of exemplars. For each exemplar pair, the Fisher linear discriminant is fitted based on the data from one of the splits. The data from the other split is then projected onto the discriminant line. The linear discriminant *t* value (LD-*t*, Kriegeskorte et al., 2007, Nili et al, 2014, Walther et al., 2015) would then be obtained by computing the *t* value for data from that pair after projection onto the discriminant line. Under the null hypothesis, the LD-*t* is symmetrically *t*-distributed around zero. Therefore it is possible to make inference on mean discriminabilities across many pairs of stimuli by performing either fixed effects analysis or across-subjects random-effects tests (see section *** for a detailed discussion on this). The LD-*t* can be interpreted as a cross-validated and noise-normalized Mahalanobis distance (Nili et al., 2014, Walther et al., 2015). The average *t* value of all exemplar pairs would then be a measure of the discriminability of all exemplars.

The discriminabilities for any two pairs of exemplars are not independent (e.g. discriminabilities of the same exemplar with two different ones). The standard error of the average *t* value is therefore smaller than 1 (the standard error of a standard *t* value), but larger than $1/\sqrt{n}$ where n is the number of all averaged *t* values (one for each pair). Therefore we understand that treating the average LD-*t* as a proper t value would be conservative (In fact the LD-*t* values are already averaged across the folds of crossvalidation, making such a test even more conservative.).

One could alternatively use the average linear discriminant contrast (LDC), which is also known as the *crossnobis* estimator (squared cross-validated Mahalanobis distance), as a measure of exemplar disciminability (Nili et al., 2014; Walther et al., 2015).

### 2.2.3 EDI or average LD-*t* after removing the effect of univariate activation from the dissimilarity measures

Response-pattern dissimilarities could be influenced by differences in the average activation of a brain region. Removing the contribution of the activation differences might be interesting in scenarios where only pattern differences are investigated.

The correlation distance for any pair of conditions is 1 minus the inner product of the standardized activity patterns after subtracting the regional average activation from each. To remove the effect of univariate activation from the Euclidean and Mahalanobis distance, we computed the distance measure after subtracting the across-response-channels mean from the pattern of each condition (results from these are denoted as "average activation removed" in Figures 8 and 10)

$$\tilde{\mathbf{b}}_i = \mathbf{b}_i - \overline{b}_i \mathbf{1}$$

where $\overline{b}_i$ is the average value of all response channels for condition i and **1** is a 1 by number of response channels row vector containing only ones. The LD-*t* is computed as explained in Nili et al., (2014). However, after estimating the discriminant weights from the training data, we subtract the component that is along **1** from the weight vector. The reasoning is that the component along **1** corresponds to the overall differences in activation for the two conditions, hence removing it from from the weights will result in a LD-*t* value with no contribution from activation differences. More specifically, we estimate the weight vector, w, from the training data (e.g., dataset 1) and replace it with $\mathbf{w} - \langle \mathbf{w}, \mathbf{1} \rangle$, where $\langle \ \rangle$ denotes the inner product.

A post-hoc way to assess the contribution of activation differences to EDI would be to compute a split-data RDM based on the absolute value of the activation differences. Now, if EDIs obtained from the univariate sdRDMs are positive, then exemplars are decodable on the basis of regional activations alone. Furthermore if the univariate EDIs are less than the original EDIs, it implies that the decodability in the multivariate response space is above and beyond from decodability based on the regional mean-activation.

## 2.3 Tests for inference

### 2.3.1 Subject as random effect

Treating subjects as a random effect allows inference about the population from which the subjects were randomly drawn. For random-effects analysis of EDI values, the summary statistic is first estimated in individual subjects. Estimates of all subjects are then jointly tested using either a one-tailed *t* test or a one-tailed Wilcoxon signed-rank test.

#### 2.3.1.1 One-sided *t* test

The standard method for testing EDIs is the *t* test. Since there is a hypothesis about the direction of the EDI (i.e. positive EDI for exemplar information), a one-tailed test is appropriate. The *t* test assumes that the data come from a population that is normally distributed under the null hypothesis ($H_0$).

#### 2.3.1.2 One-sided Wilcoxon signed-rank test

In cases where the assumption of normality of the data seems unreasonable, one might consider using non-parametric alternatives to the *t* test. In this paper, we consider the Wilcoxon signed-rank test (Wilcoxon, 1945) as the non-parametric alternative. This non-parametric statistical test compares the median of its input to zero. In this test the EDIs are first ranked according to their absolute values. The difference of the ranks for the positive and negative EDIs is then computed. The p-value corresponding to the difference of the ranks is the output of the test.

Another valid nonparametric test is the sign-test (Dixon & Mood, 1946). Significance of the sign-test is merely based on the number of positive and negative values in a given sample. (given the number of positive and negative samples, the p-value would be directly computed from the Bernouli distribution). Therefore the sign-test discards any information about the magnitude of the samples and is not considered an appropriate non-parametric alternative for *t test* in testing EDIs.

### 2.3.1.3 The *t* test and Wilcoxon signed-rank test are affected by different properties of the EDI distribution

*t* test and Wilcoxon signed-rank test base their inference on different statistics of the data distribution: while the former computes the mean and infers the standard error around it based on the sample variance, the latter computes a tail probability that relates to the sum of positive ranks of the data. These characteristics are independent from one another, therefore the tests may yield considerably different p-values depending on the degree to which each feature is pronounced in the data. To illustrate this, we simulated four sets of 12 EDIs, each by drawing random data points from four Gaussian distributions with different means and variances (Fig. 3). We then submitted each set to a *t* test and a Wilcoxon signed-ranked test (both right-tailed and testing against the null hypothesis that the data come from a zero-mean distribution) and thresholded the resulting p-values by the conventional p<0.05 criterion.

In set A, the mean of the sample is above zero and the sample variance is small; therefore, both tests yield a significant p-value. In set B, the sample mean is more positive than in set A. However, although all EDIs are positive, an outlier in the sample (value close to 30) drastically inflates the sample variance. This leads to a non-significant p-value when applying the *t* test, because the outlier increases the standard error and therefore diminishes the *t* value. By contrast, the Wilcoxon signed-rank test is not drastically affected by the outlying EDI and yields a significant p-value. The reason for this is that while the outlier has the highest rank, the rank does not contain any information on how far away this value is from the sample mean. The signed-rank test is therefore more robust against extreme outliers than the *t* test.

In set C, the variance is larger than in set A, but deviations from the sample mean are still relatively small. Therefore, the *t* test returns a significant p-value, indicating the mean EDI of the sample is different from zero, suggesting exemplar information. On the contrary, submitting the same set of data points to the Wilcoxon signed-rank test yields a p-value that is above the significance threshold. The reason for this is that a substantial proportion of EDIs in the sample are negative, making the tail probability too small to result in a significant p-value. In this case the *t* test is the more sensitive test because it accounts for the overall sample variance rather than ranks. Finally, in set D the mean is close to zero, hence both tests do not reject the null hypothesis.

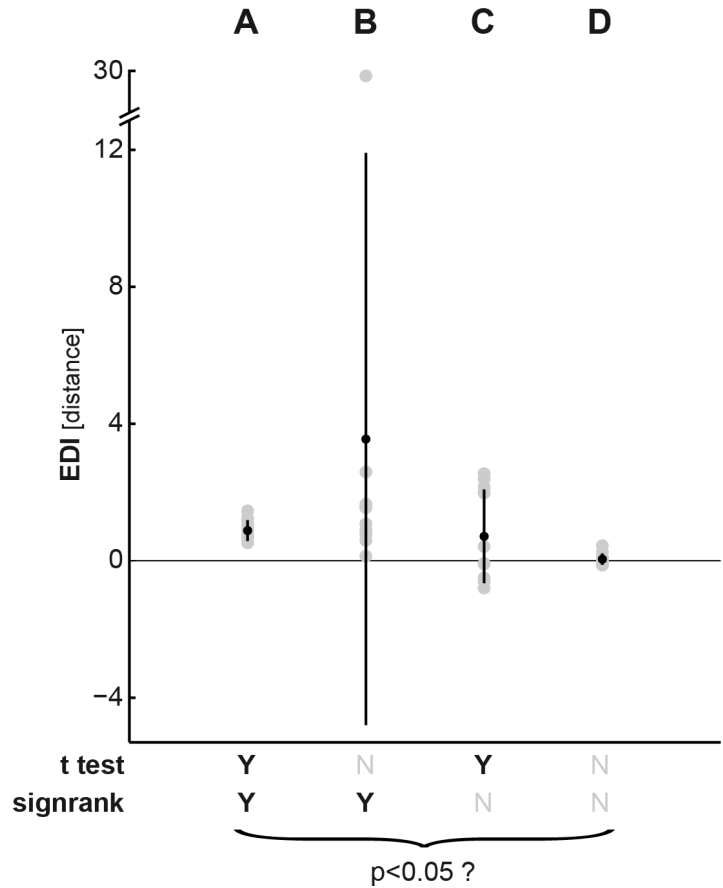

**Figure 3: *t* test and Wilcoxon signed-rank tests examine different null hypotheses about the data distribution.** The graph depicts four different sets of simulated EDIs (A to D) sampled from Gaussian distributions with different means and variances. Gray points indicate the sample data. The superimposed black point depicts the mean of each sample. The sample variance around its mean is depicted by black lines emanating from the mean (i.e., the black dot). Each set consists of 12 values, representing 12 subjects, which are submitted to both a *t* test and a Wilcoxon signed-ranked test. Both tests are right-tailed (meaning the tail extending into the positive direction) and test the null hypothesis that the data come from a distribution with zero mean (meaning no exemplar information across participants). p-values were pronounced significant (Y) if they passed the conventional threshold of $p<0.05$. Otherwise, they were non-significant (N). (A) The sample mean is positive and the sample variance is small, hence both tests pass the significance threshold. (B) The sample mean is positive. However, an extreme value in the sample inflates the variance. Therefore the *t* test does not return a significant p-value anymore. (C) The mean is positive, but the sample contains a considerable number of negative EDIs. Accounting for the weight of the ranks of the data points, the Wilcoxon-sigend rank test penalizes this presence of negative data points, therefore its p-value does not pass the significance threshold. (D) The sample mean is close to zero, hence neither test indicates significant exemplar information.

### 2.3.2 Single-subject or subject as fixed effect

In clinical cases or basic research, it is sometimes desirable to perform tests at the single-subject or group-average level. For example we may want to test if familiar faces are represented distinctly in one subject or in a particular group (e.g. patients). These tests do not require generalization to the larger population from which the current subjects are sampled from. We suggest two tests for the fixed effect analysis of exemplar information.

### 2.3.2.1 RDM-level condition-label randomisation test for EDI

Under the null hypothesis, the within-exemplar and between-exemplar dissimilarities are exchangeable. Therefore for a given split-data RDM (single-subject sdRDM or group-average sdRDM, for single-subject or group-level fixed-effect analysis, respectively), one can independently permute the rows or the columns of the sdRDM many times and compute the EDI at each iteration. The p-value for the non-parametric test is then the proportion of more extreme (greater) values in the null distribution compared to the EDI for the single subject or group average. So if N denotes the null distribution of EDIs and $EDI_m$ the single-subject or group-average EDI, the p-value is obtained according to the following formula:

Eq. 3
$$p = \frac{n(N > EDI_m)}{n(N)}$$

where *n()* is an operator that counts the number of elements in a set (i.e. its cardinality).

This test can be efficiently implemented by using the mean of the diagonal entries as the test statistic. Since the sum of the diagonal and off-diagonal entries is constant across permutations, one can equivalently compute the p-value as the proportion of diagonal averages (across many permutations) that are smaller than the diagonal average for the single subject or group.

### 2.3.2.2 Pattern-level condition-label randomization test for average LD-t

In contrast to sdRDMs, the within-exemplar distances are not estimated in LD-*t* RDMs and the exchangeability (under $H_0$) could not be applied in the same way. However, one can still estimate the null distribution of the EDI by computing the LD-*t* RDMs under the null hypothesis for one subject or group of subjects. For single-subject inference, we fit the discriminant line for the training dataset (*e.g.* dataset 1) and test it on the remaining data (*e.g.* dataset 2) after condition-label randomisation (randomly shuffled test-data). We then obtain the null distribution of the average LD-*t* values by aggregating the results from the null LD-*t* RDMs (*i.e.* LD-*t* RDMs obtained under the null hypothesis for different iterations). Fixed-effect group-level analysis is carried out in the same way as single-subject analysis. The null LD-*t* values are estimated for each subject and the null distribution for group-level analysis will be the distribution of subject-averaged values. Note that this test requires more computations than the RDM-level condition label randomisation test.

## 2.4 Scenarios for assessing the statistical tests

A statistical test can lead to errors in two cases: 1) When the data comes from the null distribution and the test gives significant results (false positives, *type I error*). 2) when the alternative hypothesis holds but the test does not detect it *(false negatives, type II error)*. If the type I error is large, the test lacks *specificity* and is not valid. If the type II error is large, the test lacks *sensitivity* and is not powerful. Ideally one would seek a test that is both sensitive and specific. In this section, we first test the validity of the underlying assumptions for the *t* test of EDIs and then investigate the specificity and sensitivity of different EDI tests.

### 2.4.1 H0: simulation

Simulations allow us to simulate multivariate ensemble vectors with known properties. We use a number of parameters to simulate multi-normal activity patterns under H0 for two datasets in each simulated subject. For any point in the parameter space (*i.e.* any combination of parameters), we simulated patterns for a large number of subjects (10,000 subjects) for each dataset. The distribution of the EDIs is then a good estimate of the EDI null distribution for that particular set of parameter values. The main purpose of this rich estimation of the EDI null distribution is to assess the validity of the required assumptions for the *t* test. For the conventional *t* test approach to give interpretable results, the null distribution of the population needs to be zero-centered and reasonably Gaussian.

Figure 4 illustrates our simulation setup. For any combination of parameters, we apply three different tests to the simulated null EDIs:

- *One-sided Wilcoxon signed-rank test*: Tests if the median of the EDI null distribution is different from zero
- *Lilliefors test of Gaussianity*: Tests if the EDI null distribution is Gaussian. The Lilliefors test (Lilliefors, 1967) is a normality test with its $H_0$ being that the data is normal (unknown mean and standard deviation). This test is based on the maximum discrepancy between the empirical distribution function and the cumulative distribution function of the normal distribution with the estimated mean and estimated variance.

- *One-sided t test*: Tests if the mean of the EDI null distribution is different from zero. Considering this allows us to estimate the false-positive rate of the standard *t* test approach for every point of the parameter space.

As illustrated in Figure 4, the simulated multivariate responses were based on five parameters. All simulated patterns can be imagined as vectors in a space with as many dimensions as response channels (e.g. voxels in fMRI). The *activation component* is a univariate component (emanating from the origin) that affects all response channels of all exemplars equally in both datasets. The *pattern component* is then the variance of activity in each response channel that is added to the activation component. Here, zero pattern component variance means that the centroid is on the all-ones vector, i.e. the response is equal in all response channels and equal to the grand mean, hence there is no spatial variability across the simulated response channels for the centroid. Conversely, a high variance means that the average pattern across splits and exemplars has great spatial variability. Adding activation and pattern component gave the response pattern centroid. Response patterns for two data splits of individual exemplars were then simulated by taking random samples from a Gaussian distribution centred at the centroid with a certain noise variance for each data split (denoted by the *noise component*). This noise variance parameter therefore determined the variability of the exemplars within and between repetitions. This is sensible because under the null hypothesis, all exemplars are indiscriminable and the within- and between-exemplar variabilities are equal.

The remaining two parameters are the *number of response channels* and the *number of exemplars*. Consider an fMRI experiment investigating whether the representations of a brain region distinguish between different face images. In that case, the number of exemplars would correspond to the number of face images that were presented to each participant and number of response channels would be the number of voxels whose activities were recorded during the scan and were investigated. Note that in practice the values of these two parameters are chosen by the analyst/experimenter. One can intuitively think that the number of exemplars may play an important role. While having more exemplars reduces the sampling error by having richer samples of the stimulus space, it also increases the gap between the number of diagonal and off-diagonal estimates of an sdRDM. Therefore one can speculate that more exemplars may not be as advantageous for the *t* test, because this also makes it more likely that the underlying assumptions of the test are violated. Exploring the effect of the number of response channels is also important. It would be essential to know how exemplar information in distributed patterns depends on the number of response channels. For example in fMRI analysis, researchers often replicate the same effect for a range of ROI sizes (e.g., Kriegeskorte et al., 2008b). This analysis helps reveal potential special *peculiar* effects for some number (or range of numbers) of response channels that are due to sensitivity of the tests or validity of the required assumptions and not the properties of the distributed representations.

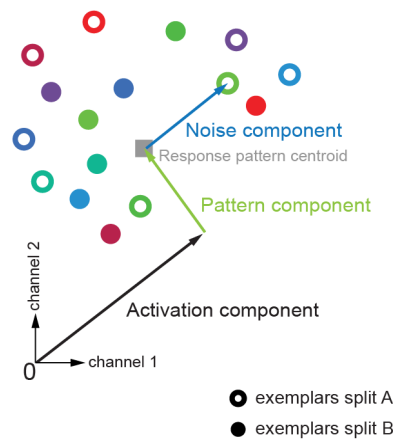

**Figure 4: H$_0$ Simulation settings.** We use simulations to test the distribution of EDIs under the null hypothesis, H$_0$ (here illustrated for only two response channels)**.** For both datasets 1 and 2, patterns of exemplars were simulated in a multi-dimensional response space. The number of dimensions of the response space is the same as the number of response channels (e.g. voxels in a region of interest in fMRI data). Response patterns were simulated by first moving along the all-ones vector to reach the average activation, whose strength was determined by the *activation component*. A Gaussian pattern with a *pattern component* variance was added to the activation component, resulting in the response pattern centroid. Individual exemplars of each data split were then generated by adding random Gaussian noise with a *noise component* variance to the centroid. At each iteration of the simulation, the same centroid underlay all exemplars in both data splits (as exemplars are indiscriminable under H$_0$). Exemplar labels were randomly assigned to experimental conditions and their repetitions. For details of the simulations see section 2.4.1.

Exploring the parameter space for multiple simulation settings can lead to false positives due to the testing of multiple hypotheses (each test is repeatedly applied for all possible combinations of the parameters). To control the type-I error rate we keep the false discovery rate (FDR) at 5% (Benjamini and Hochberg, 1995). Moreover, we also report results from applying the more conservative Bonferroni correction for multiple comparisons (i.e., controlling the family-wise error rate).

### 2.4.2 H0: simulation by shuffling fMRI data

The null hypothesis, H$_0$, can also be simulated from real fMRI data. We simulate H$_0$ from data at the single-subject level and obtain group data under H$_0$ by aggregating the simulated data from all subjects. For tests that rely on statistics obtained from an sdRDM (e.g. EDI based on Pearson correlation distance), null-sdRDMs are obtained in each subject by permuting the rows and columns independently (note that exchangeability holds for H$_0$). For tests that rely on the LD-*t* values, the exchangeability is applied by randomly permuting the order of exemplar predictors in the design matrices of both data splits. The group aggregate is estimated for a large number of $H_0$ iterations and each test is applied to the null data. Once the p-values are obtained, we compute the proportion of significant scenarios and compare it to what is expected under $H_0$ (*i.e.* number of iterations times the number of tested scenarios times the threshold). We choose the conventional threshold of 5 %.

This procedure estimates the false positive rate (type-I error rate *i.e.* proportion of an incorrect rejection of the null hypothesis amongst all tested hypotheses) and allows validating the tests without thorough exploration of the parameter space (i.e., the approach explained in 2.4.1).

### 2.4.3 H1: fMRI data

In order to assess the sensitivity of different tests, we apply all tests to the same data and a wide range of exemplar-discriminability test scenarios. To do this, fMRI data from six regions of interest (i.e., V1, V2, V3, LOC, FFA, and PPA) of 17 subjects were considered. In addition, we consider discriminability of different subsets of experimental conditions. Figure 5 shows the regions of interest and the different stimulus sets. Pattern dissimilarities were computed based on the beta coefficients from the GLM. The design matrix consisted of one regressor per exemplar and six motion parameters.

Having six ROIs and seven discrimination sets (42 scenarios in total) includes a range of effects. For a given threshold, the number of significant scenarios is an indicator of the power. A test that is more powerful will give more significant cases compared to a less powerful test. If an effect is very strong, all tests would most likely detect it (resulting in significantly small p-value) but a weak effect will only be detected by a sensitive test. Thus, the comparison is fair since different tests are applied to the same data and also reasonably general since a range of effects are considered.

Applying all tests to the same dataset would not allow statistical comparison of the power of different tests. For that, we need to estimate the "variability" in the power of tests as well. To do that, we employ subject bootstrapping (resampling subjects with replacement), repeating the same procedure for a large number of subject replacements. At each bootstrap iteration, tests are applied to the bootstrapped group-data and the number of significant scenarios are counted. The standard deviation of the bootstrapping distribution would be an estimate of the standard error of the mean for the actual data (Efron and Tibshirani, 1994). We obtain p-values for each pair of tests by computing the proportion of cases where the number of significant cases is different in the two tests (*e.g.* for two tests, $test_1$ and $test_2$, the p-value for the null hypothesis that the power of $test_1$ is greater than the power of $test_2$ is the proportion of bootstrapping iterations in which the number of significant cases reported by test$_1$ is less than or equal to the number of significant cases reported by test$_2$).

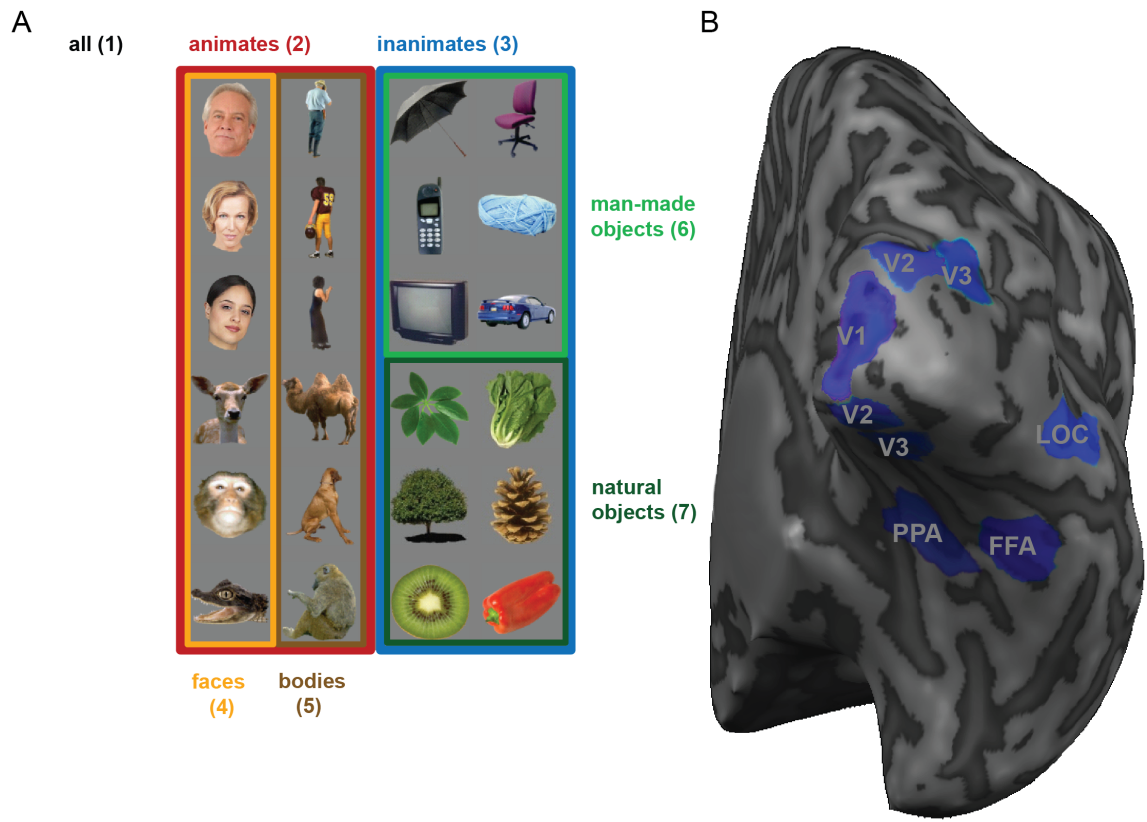

**Figure 5: Different exemplar sets (A) and ROIs (B) considered for the analysis of real fMRI data.** In order to evaluate the different tests, we apply the tests to real fMRI data. Having data from different regions of interest (V1, V2, V3, LOC, FFA and PPA, displayed for an example subject on the right) and testing the discriminability for different exemplar sets (*i.e.* different sets of experimental conditions, displayed on the left) confronts the tests with a wide range of situations.. To obtain an estimate of the type I error, different tests were applied to randomised fMRI data (simulating the null hypothesis with real data). Seven subsets of conditions are considered: 1) all 24 images 2) 12 images belonging to the animate category 3) 12 images belonging to the inanimate category 4) 6 face exemplars 5) 6 images of bodies 6) 6 man-made objects 7) 6 natural objects.

# 3. Results

## 3.1 Simulation results: all tests are empirically valid

The simulations enabled us to test hypotheses about the EDI distribution under H$_0$. In particular, we explored a wide range of parameter settings (*e.g.* number of exemplars, *etc.*). For each setting, we assessed the validity of the *t* testing approach by testing whether the null distribution conforms to the *t* test distributional assumptions and if the *t* test protects against false positives at a reasonable rate. Furthermore, we simulated $H_0$ using real fMRI data and obtained estimates of the false-positive rate for all exemplar discriminability tests (explained in section 2.3). In both cases (simulating $H_0$ using real or simulated data), if the tests are valid, the false-positive rates should not be different from what is expected to be significant by chance (the number of false-positives depends on the threshold and the number of tests, e.g. when we apply a threshold of 5 % to 100 tests under $H_0$, we expect 5 tests to be significant).

### 3.1.1 The *t* test assumptions are not met: the EDI is zero-mean, but not Gaussian under $H_0$

Although the EDI is not exactly symmetrically distributed about 0 under $H_0$, in practice it comes very close to a symmetric distribution about 0, for three reasons. (1) Though the representational distances are positively biased and have an asymmetric distribution, their distribution becomes more and more symmetrical as the dimensionality of the patterns (i.e. the number of response channels, e.g. voxels) increases. Even for very small numbers of response channels (e.g. five) the distances approximate a symmetrical distribution under $H_0$. (2) When dissimilarities are averaged to obtain diagonal and off-diagonal means, the distribution of each of these means even more closely approximates symmetry as it becomes more Gaussian (according to the central limit theorem). (3) The variances of the diagonal and off-diagonal means are clearly different, but this difference is smaller than one might intuitively expect. The diagonal elements are independent, so their average has a variance reduced by factor $\sqrt{N}$. The off-diagonal elements are dependent and although $N^2$-N of them are averaged, the reduction in the variance is much smaller than factor $\sqrt{N^2 - N}$.

For these reasons, it is very difficult indeed to find a scenario by simulation where the EDI is not approximately symmetrically distributed about 0 under $H_0$. Figure 6 gives an example of a simulated extreme scenario, which we selected to illustrate the theoretical violations of the normality and symmetry assumptions. The upper two panels illustrate the approximate symmetry of the dissimilarities and their diagonal and off-diagonal means. The lower panel shows the EDI distribution, which is also close to, but not exactly, zero-mean symmetric. The EDI in this selected simulated null scenario is slightly but significantly non-Gaussian (Lilliefors test) and the *t* test and Wilcoxon signed-rank test are also significant.

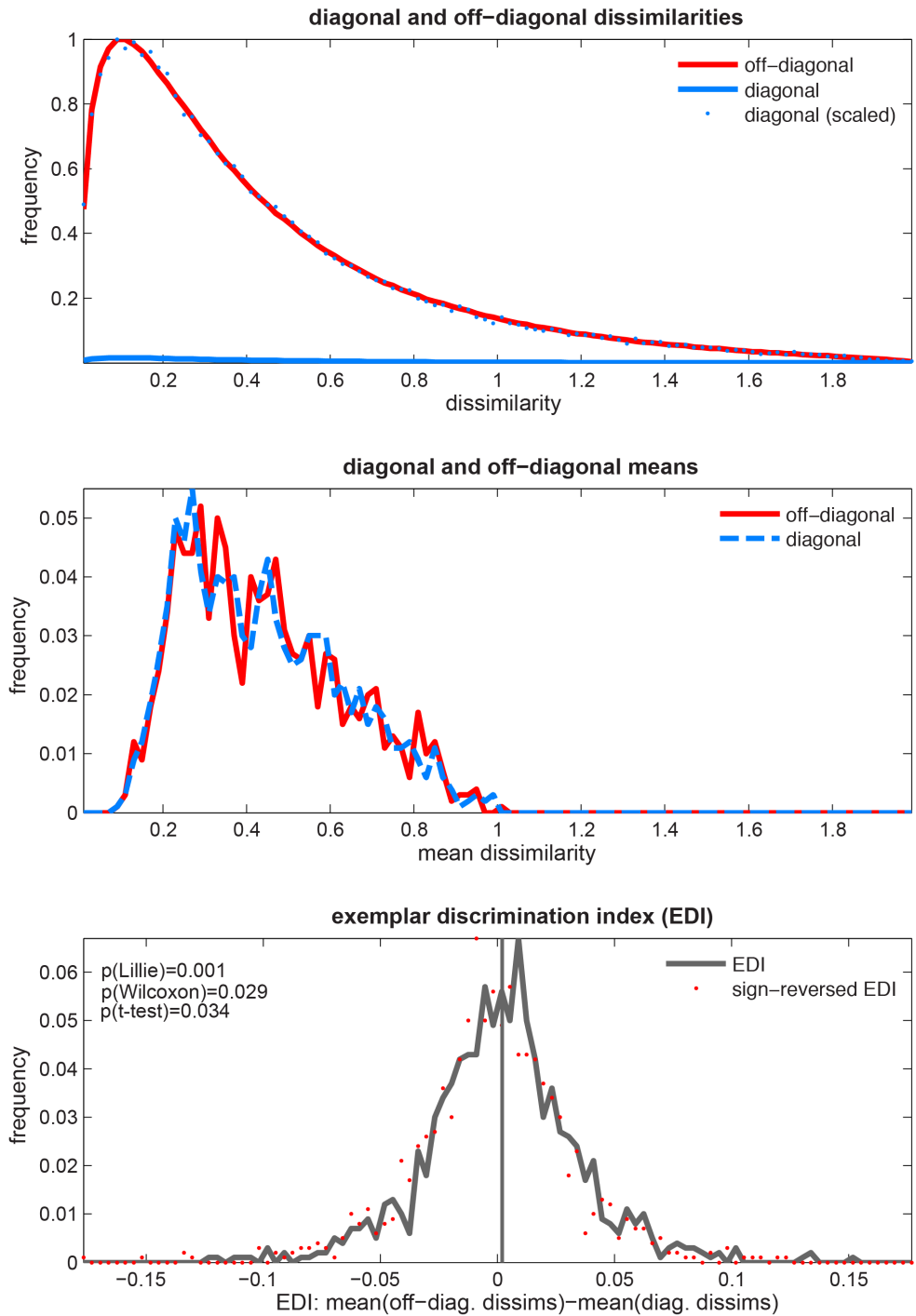

**Figure 6: The EDI can be non-Gaussian under the null hypothesis (H$_0$).** *Top panel:* The distribution of the diagonal (blue) and off-diagonal (red) dissimilarities. Distributions are obtained from the dissimilarities aggregated across pairs and simulated subjects. There are fewer diagonal entries but they are sampled from the same distribution as off-diagonal dissimilarities under the null hypothesis. Therefore the two distributions differ only by a scaling factor. *Middle panel:* The distribution of the diagonal and off-diagonal average dissimilarities. Data were simulated for each subject (1,000 in total) and the distributions were obtained by pooling the averages across all subjects. *Bottom panel:* The EDI distribution in this case is significantly non-Gaussian. Furthermore, a one-sided *t* test or Wilcoxon signed rank test both yield significant p-values. This means that EDI is not strictly zero-mean and Gaussian under H$_0$. This could potentially inflate both types of error and weaken inference based on the *t* test. For each of the 1,000 subjects, data were simulated for 64 exemplars and five response channels.

### 3.1.2 The *t* test is empirically valid because of its robustness to violations of its assumptions

In the previous section we showed that the EDI is not strictly Gaussian under $H_0$. Here, we used simulations to see how often these violations occur. To this end, sdRDMs were computed under the null hypothesis for a large number of simulated subjects. Simulations were carried out independently for every point of the 5-dimensional parameter space. The parameters were the number of exemplars, number of response channels (*e.g.* voxels) and three others characterizing the multivariate response space (see Fig. 4 and section 2.4.1). At each point of the parameter space, we tested if the EDI null-distribution was Gaussian and if it was zero-centered (the two main assumptions of the *t* test of EDIs). Additionally, for every point, we estimated the false positive rate of the *t* test. Estimating the false positive rate at each point in the parameter space was carried out by independently repeating the whole simulation 1,000 times and then obtaining the false positive rates from the proportion of cases where the *t* test is significant for a specified threshold.

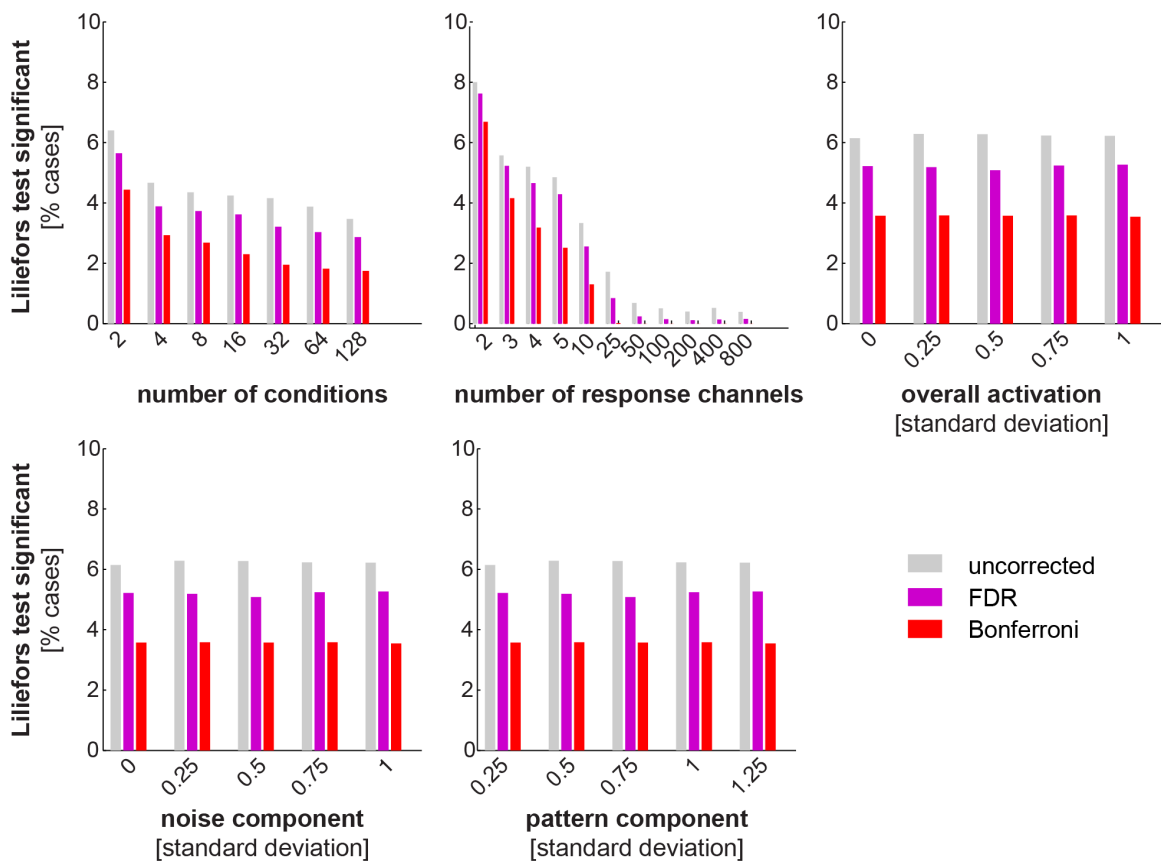

**Figure 7: Marginal histograms of deviations from Gaussianity of the simulated EDI null distribution.** In order to richly characterise the properties of the EDI distribution under H0, we simulate data for a large number of subjects for a range of values of any parameter. At every point of the cross-product parameter space, sdRDMs (based on Pearson correlation distance) were obtained from simulated patterns for 10,000 subjects (patterns were simulated independently for each subject). Lilliefors tests (testing whether the distribution of EDIs is significantly non-Gaussian) were applied to the EDI distributions. These bars correspond to results from the Lilliefors Gaussianity test. A significance threshold of 5 % was applied to the tests. Each column also corresponds to one dimension of the parameter space. The first two (number of exemplars and number of response channels) mainly depend on the experiment and analysis and the other three characterise the multivariate response space. For example, the red bar on the leftmost bar-graph would correspond to the proportion of all Lilliefors tests that were significant ($p < 0.05$, Bonferroni correction) when the number of conditions were fixed at 2, 4, 8, etc. and the other parameters had any possible value. Two other tests were performed in a similar way to test EDIs for the simulated data under H0. One was a one-sided Wilcoxon signed rank test and the other a one-sided *t* test. The signed rank test was applied to assess whether the EDIs were zero centered and the *t* test was applied to obtain an estimate of the false-positive rates. Interestingly, for both tests less than 1% violations occurred ($p < 0.05$) and none of those survived correction for multiple comparisons (FDR and Bonferroni, $p < 0.05$).

Figure 7 shows the results for the Lilliefors test. Each panel corresponds to one of the parameters of the parameter space (5 in total, see section 2.3.1). Each bar graph gives the marginal histograms for the frequency of violations of the Lilliefors test for different levels of a parameter. Frequency of violations for the other two tests, i.e. *t* test and signed-rank test were close to zero for the different levels of all parameters (after controlling FDR at 5%) and not displayed. As expected, the null distribution is very rarely centered at a point different from zero.

Importantly, in accordance with the theoretical argument for the non-Gaussianity of the EDI, cases in which the EDI null distribution was significantly non-Gaussian were not infrequent. The validity of the assumptions also seemed to depend on the parameter level (e.g. fewer response channels are more likely to give rise to a non-Gaussian EDI null distribution). Interestingly, despite these violations of assumptions, the false-positive rates of the *t* test were not significantly inflated in any of the tested scenarios (corrected for multiple comparisons). These seemingly contradictory results can be explained by the fact that the *t* test is robust to violations of its assumptions. Overall, these results suggest that the assumptions are violated in some cases, but the violations are small and do not significantly inflate the false positives rate of the *t* test.

### 3.1.3 All tests of exemplar information protect against false positives at the expected rate.

Our simulations showed that the one-sided *t* test is a valid approach to test EDIs. Section 2.2 listed a variety of EDI tests including the commonly used *t* test. In this section we estimated the false positive rate (type-I error rate) of all exemplar information tests. To this end, we applied the tests to many instantiations of group data under $H_0$. Each instantiation was obtained by shuffling fMRI data (as explained in 2.4.2). Fig. 8 shows the false-positives rates of the different tests.

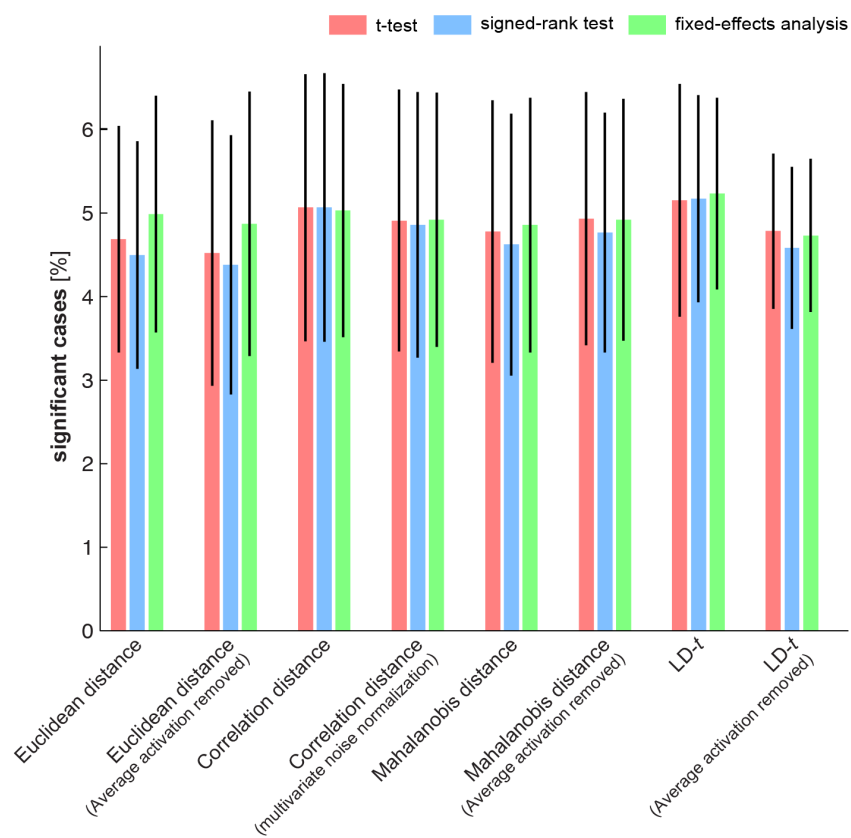

**Figure 8: False-positives rates for different tests of exemplar discriminability.** We estimated the false positive rate (type-I error rate) by applying tests to randomised fMRI data (simulating $H_0$, see section 2.4.2 for details). The null data were simulated for a large number of iterations. At each iteration an ROI was randomly selected and data for each subject from that ROI were randomised by permuting the condition labels for a full set containing experimental conditions from both data splits. Randomisation was independently carried out for the two data splits. Under the null hypothesis, exchangeability implies that there should not be a pattern difference between replications of the same exemplar and a different exemplar. At each iteration (1,000 in total), we simulated group level data under $H_0$ and applied all tests to it. We then computed the average percentage of false positives (proportion of significant cases from the total number of 1,000 simulated cases, i.e. height of the bars) and their standard deviation (error bars). All tests had a false positive rate that was not significantly different from 5% after a threshold of 0.05 was applied to the tests ($p > 0.5$, two-sided Wilcoxon test). This means that all tests protect against false positives at an expected rate and are therefore valid.

As the results show, all tests have an acceptable false positive error rate: at a significance threshold of 5%, about 5% of the tests give significant results.

Simulating the null hypothesis using real data is more realistic than using simulations, as it incorporates all the complexities of measured data. In simulations, it can be impractical to model voxel dependencies or the extent to which response patterns change from one session to the other (different data splits). Using real data inherits all those dependencies and allows answering the same questions. However, one disadvantage of real data is that parameters of interest cannot be studied as principled as in simulations and that the conclusions may not hold for other types of neuroimaging data. In this case the consistentcy between the results from the simulated and real data makes us confident that all the propsed tests are valid and can be used to test exemplar information from brain measurements.

It must be noted that having a reasonable false positive rate is a necessary criterion for the validity of a test and if any test gives greater false positive rate than what is expected by chance, the test would not be considered further.

### 3.1.4 The EDI and linear decoders are sensitive to variance differences between conditions, whereas crossvalidated distance estimators are not

The multivariate response to each experimental condition can be treated as a sample from a high-dimensional probability distribution. The condition-related distributional differences may be of neuroscientific interest. For example, exemplar information may be present in higher-order activity-pattern statistics. Testing for mean differences can miss those effects. If two condition-related pattern distributions have identical means and different variances (Fig. 9), then there is mutual information between the response pattern and the experimental condition, which the brain might exploit.

Figure 9A illustrates that the EDI is sensitive to differences in the condition-related variances when the pattern means are identical. We consider two conditions A and B sharing the same mean pattern. The pattern-estimate distribution for A has small variance and that for B has large variance. To complement the visual intuition provided by Fig. 9 with a simple numerical example, consider the extreme case where the variance is zero for condition A. The Euclidean distance between samples **a** (from A) and **b** (from B) is the root mean square (across voxels) of the difference **b-a**, so we can use the standard deviations (normalized root mean squares) of the distances with equivalent results (down to the proportionality factor). The expected distance between pattern estimates from A and B here is $\sqrt{(var(A) + var(B))} = \sqrt{0+1} = 1$. The expected distance between different pattern estimates from A is $\sqrt{var(A) + var(A)} = 0$. The expected distance between different pattern estimates from B is $\sqrt{var(B) + var(B)} = \sqrt{2}$ So the expected EDI is $1 - (0 + \sqrt{2})/2$ and larger than zero.

Figure 9B illustrates that a linear decoder is also sensitive to differences in the condition-related variances (Hebart & Baker, 2017). Placing the decision boundary to one side of the smaller-variance distribution can yield above-chance decoding accuracy. Note that our illustration here uses isotropic distributions. However, similar results obtain for nonisotropic distributions, and spatial whitening does not resolve this issue.

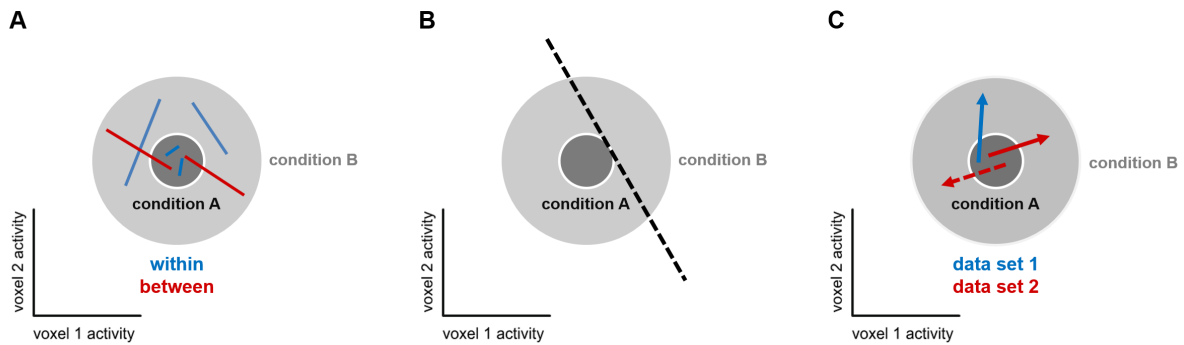

**Figure 9: EDI and linear decoding accuracy are sensitive to differences in variance between conditions, but crossvalidated distance estimators are not.** Consider two distributions of pattern estimates for two conditions (A, dark gray; B, light gray), which have identical means (central point), but different variances (small for A, large for B). (**A**) EDI: The mean distance between conditions (red) is larger than the mean distance within conditions (blue), because the distances within condition A are very small. The EDI, therefore, is sensitive to the difference between the conditions. (**B**) Linear decoder: A linear decoder may maximise its accuracy by placing the decision boundary to one side of the condition-A distribution. This way the decoder gets 100% accuracy for condition A. For condition B, accuracy is below chance (50%), but above 0. On average across conditions, the accuracy will be above chance (50%). Linear decoders, therefore, are sensitive to differences in variance (like the EDI). (**C**) Crossvalidated distance estimator: A crossvalidated distance estimator (e.g. crossnobis/LDC or LD-*t*) is a mean of inner products of pattern-difference vectors (arrows: condition-B pattern estimate minus condition-A pattern estimate). On each fold of crossvalidation, an inner product is computed between a training-set vector (blue) and a test-set vector (red). When the distributions are point-symmetric about the mean (e.g. Gaussian), each test-set vector (solid-line red arrow) has an equally probable opposite twin (dashed-line red arrow). The inner products of the training-set vector with the twin test-set vectors (red) will have opposite sign and the same absolute value. These pairs of values will cancel in the expectation. Crossvalidated distance estimates are therefore insensitive to variance differences (as long as the distributions of the pattern estimates are point-symmertic, e.g. non-isotropic Gaussian, for both conditions).

Although the brain might exploit variance differences using linear or nonlinear readout mechanisms, these are more difficult to interpret than mean differences. Moreover, in designs where conditions are not balanced (e.g. different numbers of repetitions for different conditions), the pattern estimates will differ in their variance even if there is no information about the exemplars in the responses at all. In this scenario, the variance differences do not reflect neuronal information that the brain might exploit, but arise as an artefact of the experimental design and analysis (with the researcher injecting the information about the conditions by using the design matrix). The EDI and linear decoders are therefore not appropriate measures of pattern discriminability for unbalanced designs.

We may prefer more conservative measures of pattern discriminability, which are sensitive to differences in pattern mean, but not to differences in pattern variance. Crossvalidated distance estimators, such as the LDC (also known as the crossnobis estimator) or the LD-*t* (Kriegeskorte et al., 2007; Nili et al., 2014; Walther et al., 2016) provide sensitivity to mean differences, but not to variance differences, as illustrated in Fig. 9C. Crossvalidated distance estimators do not use a threshold but estimate the distance as a mean of inner products. As long as the condition-related distributions are point-symmetric, these estimators are zero in expectation (unbiased) when the pattern means are identical. This still holds when the distributions differ (in variance and/or shape) between the two conditions. In particular, the distributions could be nonisotropic Gaussian distributions with different covariance matrices. Having more data (e.g. more independent measurements) would bring the estimates closer to the true values and can always improve the precision.

### 3.1.5 EDI versus exemplar decoding accuracy

In addition to obtaining an EDI, split-data RDMs could also be used to obtain an estimate of exemplar decoding accuracy. A minimum-distance classifier would be able to successfully decode exemplars if the nearest neighbour of the response to each exemplar is also from the same exemplar (i.e. its replicate). Therefore, for two exemplars and two datasets, each within-exemplar distance needs to be less than the two between-exemplar distances (4 inequalities) and the decoding accuracy would be the percentage of inequalities that are satisfied. To obtain the exemplar decoding accuracy for a set of exemplars, each diagonal element of a sdRDM (N elements) would be compared to each corresponding off-diagonal element (2*(N-1) elements) and the total number of satisfied inequalities would be counted. Normalising this count by the total number of inequalities, i.e. 2N(N-1) gives an estimate of the exemplar decoding accuracy for an exemplar set. The exemplar decoding accuracy is expected to be less sensitive than the EDI because of the loss of information incurred by the counting (Walther et al. 2016). Moreover, decoding accuracy saturates at 100%, whereas the EDI continuously measures the separation of the exemplars. For these reasons, continuous nonsaturating meaures, including the EDI and pair-averaged crossvalidated distance estimates, appear preferable to exemplar decoding accuracy.

Another theoretical difference between EDI and decoding accuracy (obtained from linear decoders) is that EDI does not only capture linear information. For example, the EDI can be positive in cases where a linear classifier cannot successfully decode two exemplars. Similar to linear decoders and unlike EDI, average LDt or LDC also quantify the linear separability of representations.

### 3.1.6 Interpretational caveat of random-effects analysis for EDI

Allefeld and colleagues (2016) have argued that testing classification accuracies to chance level through the random-effects analysis implemented by a t-test does not provide population inference. The reasoning is that the true level of the measures can never be below chance level. The same argument can be applied to EDI values. The true EDI, i.e. the EDI estimated from patterns without noise, will always be positive: the diagonals would all be zero and off-diagonal entries are also positive. Therefore, the null hypothesis that there is no exemplar information would translate to the average EDI being zero. Now, the average EDI for the population can only be zero if the EDI is zero in every single subject, effectively the null hypothesis for the fixed-effect test of the average EDI. Therefore, the random and fixed effects would converge in this case. The random-effect tests we suggest takes into account the between-subject variability and would be more conservative than the tests that treat subjects as fixed effect. All together, the across-subject EDI tests are valid tests that are subject to an interpretational caveat.

One way to circumvent that would be to convert EDI estimates to classification accuracies and test for information prevalence (Allefeld et al, 2016). However, as noted earlier, converting EDI to classification accuracies brings the undesirable property that the measure will get saturated and cannot get larger for fully discriminable patterns (Walther et al, 2015).

### 3.1.7 separating differences in overall activations and differences in patterns

Generally, EDI tests exploit all the information present in regional responses. This also includes overall activation differences. In section 2.2.3 we suggest various ways of removing the overall activation effect when computing discriminability measures. Another way to quantify the contribution of regional mean activation effects on exemplar discriminability would be to test if the exemplars could be decoded on the basis of the global response alone. This could be achived by testing EDIs from activation-based sdRDMs. In that case entries of the sdRDM would be the absolute value of the differences in the voxel-averaged regional response for same or different exemplars in two datasets. A significant univariate EDI would imply that the exemplar discriminability is at least partly driven by overall activation differences. This contribution could be removed by regressing out the activation-based sdRDM from the original sdRDM.

## 3.2 Real-fMRI-data results: Significantly greater sensitivity through multivariate noise normalization

To complete our assessment of exemplar information tests, we compared their sensitivity, i.e. the power of the tests to detect an effect when present. One way to proceed would be to simulate data with known ground truth (i.e. exemplar discriminabilities), apply the tests to the simulated data and compare them (or their ranks) in terms of their significances. However, this approach is both computationally expensive and unrealistic. In the parameter space, each parameter can span an enormous range of values. Therefore, it would be impractical to carry out a comparison for each possible combination of parameters. Moreover, as mentioned earlier, voxel dependencies and appropriate levels of replicability observed in real fMRI data are not known a priori. For these reasons we compared the tests by applying all of them to the same dataset. The dataset contained six brain regions and seven exemplar subsets (see section 2.4.3 and Fig. 5). Therefore the total number of significant scenarios for a given test could vary from 0 to 42. Note that in order to make the test comparisons valid and fair, the same significance threshold was applied in all tests. Figure 10 gives an estimate of the sensitivity of different tests based on the real fMRI dataset.

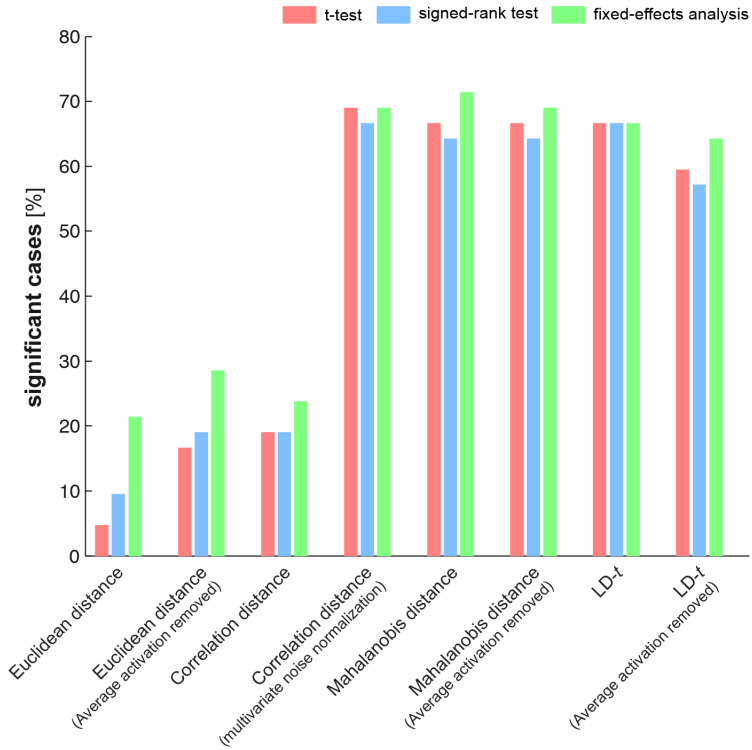

**Figure 10: Comparing the sensitivity of different tests.** Each test was applied to fMRI data from six different ROIs. In each ROI, discriminabilities of seven category subsets were assessed. A test was pronounced significant if p < 0.001. The height of the bars shows the percentage of significant cases out of the 42 tested scenarios (e.g. 50% corresponds to 21 significant tests). Tests on EDIs computed from multivariately noise-normalized response patterns were significantly more powerful (the last five triple of bars). Inference was done using subject bootstrap. All pairwise comparisons between the two groups of tests (i.e. the first three triple of bars and the next five triple of bars corresponding to unnormalized and multivariate noise-normalized data) were significant after controlling the expected false discovery rate at 0.01.

The results show a striking effect of multivariate noise normalization (Equation 1) in the sensitivity of EDI tests. All tests that are applied to data after multivariate noise normalization yielded a greater proportion of significant cases. Statistical comparison of the tests was carried out by bootstrapping subjects and applying the tests to the group data at each bootstrapping iteration. Obtaining p-values for comparing pairs of tests and controlling the false discovery rate at 1%, we concluded that the difference between the two groups of tests was significant. In other words, summary statistics that were obtained from data after accounting for the noise covariance between voxels had significantly greater power in detecting exemplar effects.

## 4. Discussion

This article explores different approaches to testing subtle within-category effects in brain representations (motivated by the theoretically-invalid t-test that was common in the literature).

### 4.1 Testing for exemplar information in condition-rich designs is important for understanding brain representations

Understanding brain representations asks for characterising the capabilities of the representations. One important characteristic would be to allow discrimination between different members of a category i.e. category exemplars. One could use a classifier and test for exemplar information by calculating the performance of a classifier that is trained to distinguish between different exemplars (e.g. employing a support vector machine, Burges 1998; Misaki et al., 2010).

In contrast to the classification-based approach that requires having many repetitions of each exemplar, we aim to extract exemplar information for condition-rich designs where many exemplars are used with fewer repetitions for each. Given the limited time for measuring brain representations, there would be a trade-off between the number of tested exemplars and number of repetitions of each exemplar. Resolving this trade-off would allow higher inferential power and a stronger claim about the representational structure of a brain region. For example if we could have more exemplars *i.e.* richer sampling of the space of exemplars, inference would be based on a larger group and that would reduce the sampling error. *Condition-rich decoding* refers to quantifying and assessing exemplar information in cases where discriminability of many experimental conditions (e.g. category exemplars) are tested.

Condition-rich decoding from distributed brain representations is possible using a simple and intuitive idea: if (on average) repetitions of the same exemplar produce less pattern-change than that produced by changing the exemplar, there is exemplar information in brain representations. Since this approach averages the effects across many pairs of exemplars, presence of exemplar information does not imply that any two pairs would be discriminable but that on average there is enough information to discriminate between exemplars on the basis of brain representations. Furthermore, in order to undertstand what exemplar distinctions are *driving* a significant EDI in an ROI, one can consider follow-up tests within the same framework. For example, in the extreme scenario, distinctions between any pair of conditions can be tested at the single-subject level or the group-level by testing the LD-t (Nili et al, 2014) or the crossnobis (Walther et al, 2015). Additionally, one can consider exploring the effect for each individual subject by investigating the EDI significance at the single-subject level. Therefore, although a significant EDI does not preclude the possibility that the effect might be driven by some set of conditions or not present in all subjects, the tools we provide here allows clarification in full depth.

## 4.2 The conventional *t* test approach is theoretically problematic but practically acceptable

The conventional approach for condition-rich decoding uses *t* test on the EDIs. The *t* test is a parametric test and, for a given sample, tests if a parameter of the sample distribution, its mean, is different from zero (i.e. the mean of the Gaussian distribution obtained under the null hypothesis). Using parametric statistics can be more powerful when the required assumptions are satisfied. However, if the requirements are not met, test results are not interpretable (Nichols & Holmes, 2002). For reasons that we explain in the paper, the null distribution of the EDI is likely to violate the required assumptions of the *t* test (e.g. violation of the Gaussianity assumption;see section 3.1.1). Therefore, we speculated that *t* test would be problematic. Using simulations, we first appreciate this concern by observing cases where applying a *t* test is wrong. This was then followed by a systematic exploration of the space of possible parameters (characterizing the response space in a wide range of settings) and showing that violations of assumptions are indeed frequent but to the extent that could mostly be tolerated by the test (although the *t* test presumes distributional properties, it is tolerant to violations of the assumptions).

## 4.3 Testing exemplar information at the single-subject level or group-level with subject as fixed effect is possible using novel randomization tests

Randomization tests, and more generally non-parametric statistics, are becoming more and more popular in the univariate analysis of neuroimaging data (Nichols and Holmes, 2002). We can also use randomization tests and non-parametric statistics in the context of testing for information in brain response-patterns. For example, randomization tests have been proposed before to test for the similarity (i.e. correlation) of two RDMs (Kriegeskorte et al., 2008b, Nili et al, 2014). Similarly, in the case of testing EDIs, randomization techniques could be employed to design non-parametric tests of exemplar information.

Hitherto, it has not been possible to test for exemplar information at the single-subject level or to do group-level analysis with subject as the fixed effect. Here we introduce tests for performing fixed effect analysis. To our knowledge this is the first attempt to fixed effect analysis of exemplar information using pattern-information analysis especially in the context of condition-rich designs. Our tests are based on randomization methods that estimate the distribution of the statistic under $H_0$. For group-level analysis, these tests are more sensitive since they ignore the between-subject variability.

## 4.4 Tests that take the covariance structure of the noise into account are significantly more powerful

In the past, only few studies have attempted to do condition-rich decoding. All those studies were based on quantifying pattern-changes by computing Pearson correlation distances and using a one-sided *t* test to test the net pattern-effect (average effect of changing exemplars minus average effect of repeating an exemplar) quantified in each subject, against zero. The *t* test is widely used in psychology and neuroscience; however, its use is less motivated for testing exemplar information in distributed patterns. The main reason against using *t* test is the theoretical concern about the distribution of the test statistic (i.e., EDI) under the null hypothesis.

Having established the validity of the standard approach, we consider different ways of testing exemplar information by exploring different tests and test statistics. In particular we introduce ways of testing exemplar information at the single-subject level or fixed effects analysis at the group level. In addition to the fixed effect tests, we also consider other tests including non-parametric alternatives to the *t* test. Those assumption-free methods can be used for testing exemplar information even in the extreme cases where *t* test is not valid.

The various test statistics considered in this paper are different due to their different ways of quantifying pattern-effects. In particular different measures could be used for computing pattern dissimilarities. Therefore, the situation is as follows: there are different ways of assessing exemplar information and the net effect could be *summarized* and test*ed* in different ways. Interestingly, we see that all the possible combinations of tests and test statistics yield reasonable false positive rates. In other words, choosing any of the proposed ways of summarizing the results and any of the tests would result in acceptable specificities. However, when looking at the sensitivities, we did not find equal levels of power for the different combinations.

By contrast, we observed that within the explored tests (i.e. *t* test, Wilcoxon signed rank test and fixed effect tests) and test statistics (i.e. EDIs based on different distance measures or average LD-*t* values), the most important factor in determining the power of the test was the way to quantify and summarize the net effect, and not the test itself. All pattern-effects that were computed after multivariate noise normalisation could be detected with a greater power than those that were not. This effect of multivariate noise normalisation afforded an almost 3-fold boost of power for the analysed dataset. Therefore, these results have clear practical implications for future studies. For example, researchers could add multivariate noise normalization to the pre-processing stages of data analysis and use the Euclidean distance - which has a simple geometric interpretation – to estimate pattern dissimilarities on the pre-processed datasets. This would have the advantage of being both simple and powerful.

We have previously presented evidence that multivariate noise normalisation renders RDMs more reliable (Walther et al., 2015). This paper documents the benefit in the particular scenario of exemplar effects, and more specifically to both within-run and between-run distances.

## 4.5 Quantifying category information with the category discrimination index (CDI)

Although category information and exemplar information consider conceptually different characteristics of representational geometries, they could both be assessed using similar methods. In this paper, we focus on exemplar information. Category information could be quantified from an RDM by subtracting the average-within-category dissimilarities from the average-between-category dissimilarities *i.e.* a category discriminability index (CDI). The CDIs could then be tested in exactly the same way as EDIs. The main difference is that for CDI there is no need to have independent measurements of the same experimental conditions and split the data into two halves. One could use the whole dataset and compute an RDM from the full dataset. Using more data (i.e. the whole dataset as opposed to data-halves for a sdRDM) may likely result in more stable patterns, which would be an advantage of this approach. Moreover, with cross-validated dissimilarity estimates like the crossnobis or the LD-t, category information could be computed by averaging the between-category discriminabilities. Since those are unbiased estimates, a significantly positive average would imply category information. It must be noted that similar to exemplar tests, category decodability using CDI or classification accuracy would be sensitive to differences in the variances for the two categories; however, average between-category crossnobis estimates would only be sensitive to category-centroid differences. Therefore, using average classification accuracy or the CDI can give significant results in the absence of category-centroid differences in cases where the designs are not balanced (e.g. different number of exemplars for the two categories) and there are variance differences due to the unbalanced design.

On a similar vein to testing EDIs, one can also subtract the average within-category dissimilarities from average between-category dissimilarities for the cross-validated dissimilarities. In this case a significant result would imply category *clustering*, which is a more strict characteristic compared to linear category decodability.

Another approach to testing category information is to test the rank correlation of RDMs with a categorical RDM that assumes equal and smaller values (*e.g.* zero) for within-category dissimilarities and larger values (*e.g.* one) for between-category dissimilarities (Mur et al., 2013, Khaligh-Razavi & Kriegeskorte, 2014). We have recently proposed using Kendal's tau-a for RDM correlations in such cases where tied ranks are predicted for either between- or within-category dissimilarities (Nili et al., 2014). In cases where there are only two categories (e.g. a binary model RDM), testing rank correlations and linear correlations would be the same. However, when there are more than two levels in the discrete model RDM, testing rank correlations would be different and more lenient than linear ones. (note that for a binary model RDM the CDI for one dataset is proportional to the linear correlation of the model RDM with the data RDM)

An alternative approach is to fit categorical RDMs to brain RDMs using linear regression and test the regression coefficients against zero (Mur et al., 2013, Jozwik et al, 2015). Similar to cross-validated RDMs, category information could be also tested by testing the average-between-category classification accuracies (Cichy, Pantazis and Oliva (2014), although see Walther et al. 2015, for a comparison of continuous measures like crossnobis versus discretised measures like classification accuracies for estimating dissimilarities).

## 4.6 Relation to model testing

Representational similarity analysis enables testing computational models of information processing in the brain. This is achieved by comparing the predicted representational geometries of models with observed geometries of the brain. Computational models can also imply exemplar information but only test for a particular representational geometry, i.e. particular configuration of exemplars in the response space. Therefore testing a model is a more focused test which would be more powerful if the hypothesis is true.

Note that the EDI is maximally general in that it is sensitive to any differences between conditions. It is highly powerful when differences exist for many of the pairs, but in principle it can detect a difference between just one pair of exemplars. Note also that the EDI can detect equidistant representational geometries, to which tests of RDM models using correlation coefficients are not sensitive (since there is no variability in the predicted distances that could be explained by a model RDM).

Additionally, one could use sdRDMs (Figure 1) to test computational models. The model prediction would be a symmetric matrix that predicts zeros along the diagonal entries. This would enable us to test the model predictions of the dissimilarities at the level of a ratio scale, not just an interval scale. It would increase the power when the model predicts very similar distances among the representational patterns. (Consider the extreme case where the model predicts equidistant patterns. In this scenario the sdRDM is required, and the model test would be equivalent to the EDI if the Pearson correlation was used). More generally, quantifying brain representations by a sdRDM would allow modelling the condition-wise differences in noise levels.

Some researchers have recommended using between-run distances for RSA (Mumford et al., 2014). They have shown that in cases where there are unavoidable structures to trial order and between-subject randomisation cannot be afforded, valid distance estimates can be obtained only when activity patterns are compared in two independent datasets (e.g. different fMRI runs). Cross-run RDMs can be computed through comparing pattern estimates from one run to the average of all other runs (ideally optimally combining the patterns would be preferred to averaging them).

The importance of presentation-sequence randomisation is not limited to single-trial RDMs. Also in RDMs for which the activity patterns are based on multiple repetitions of an item, for instance, if a condition is always presented before another condition in the first run and in the reverse order in the other run, this can increase the odds of getting a negative crossnobis estimate due to the effects of the scanner drift. Therefore, in cases where the presentation sequences are not randomised across independent measurements (e.g. fMRI runs) and subjects, using between-run distances after multivariate noise normalisation might be a good comporomise. Surely, one could ensure that the design is balanced and use crossnobis for computing RDMs.

### 4.7 Choosing the appropriate summary statistic for condition-rich decoding

In this manuscript we have explored a range of summary statistics and tests for exemplar discriminability. Table 1 summarizes the characteristics of the summary statistics. We recommend using the stimulus-pair averaged crossnobis or LDt for testing mean differences and EDIs based on multivariate noise normalisation for testing effects in mean and/or variance differences.

| test statistic | multivar. noise model | sensitive to | H₀ | inference procedure | inference scope | validity (specificity) | power (sensitivity) |
|---|---|---|---|---|---|---|---|
| **exemplar discriminability index (EDI):** mean of between-exemplar distance minus mean of within-exemplar distances | no | differences in pattern distributions (incl. mean and variance) | exemplar response-patterns drawn from the same distribution | one-sided t test across subjects | subject population | usually acceptable (despite violations of assumptions) | bad |
| | | | | one-sided signed-rank test across subjects | | good | bad |
| | | | | exemplar-label randomization | subject sample | good | good |
| | yes | | | one-sided t test across subjects | subject population | usually acceptable (despite violations of assumptions) | very good |
| | | | | one-sided signed-rank test across subjects | | good | very good |
| | | | | condition-label randomization | subject sample | | excellent |
| **average of pairwise crossvalidated discriminabilities** (crossnobis, LD-*t*) | | differences in pattern means | exemplar response-patterns drawn from distributions centered on the same mean pattern | one-sided t test across subjects | subject population | | very good |
| | | | | one-sided signed-rank test across subjects | | | very good |
| | | | | exemplar-label randomization (different exemplar labels between training and test sets) | subject sample | | excellent |
| **crossvalidated MANOVA** (pattern distinctness) Allefeld & Haynes, 2014 | | | | t test, signed-rank, permutation | subject sample or population | | very good, excellent |

**Table 1:** An overall view of different methods for testing exemplar discriminability.

## 5. Conclusions

When independent measurements of the same set of exemplars are available, condition-rich decoding can be carried in different ways. Despite the theoretical concerns against applying the *t* test to the EDIs, we validated the use of the *t* test, which is the established approach for testing exemplar information. Furthermore, we explored different ways of summarizing and testing exemplar discriminabilities and introduced a novel randomization test that is assumption-free and also allows testing EDIs at the single-subject level or group level analysis with subject as fixed effect. Comparing the different methods, we conclude that it is mostly critical to compute the EDI after applying multivariate noise normalization to the response patterns. Multivariate noise normalization considers the noise variance-covariance structure of the data and enables a more reliable estimation of RDMs. This means that computing the EDIs after multivariate noise normalization or using the average LD-*t* or crossnobis will likely result in higher sensitivity to detect exemplar information in brain representations. We also explain the difference between exemplar (and category) tests that are based on average crossnobis or LD-t and those that are based on EDI or average decoding accuracies. Average crossnobis or LD-t is only sensitive to differences in pattern mean whereas EDI and decoding accuracies are also sensitive to variance differences.

**Data availability:** The data that support the findings of this study are available on request from the corresponding author.